# Overhead-Free Computation, DCFLs, and CFLs[*]

Lane A. Hemaspaandra[†]
Department of Computer Science
University of Rochester
Rochester, NY 14627, USA

Proshanto Mukherji[‡]
Department of Computer Science
University of Rochester
Rochester, NY 14627, USA

Till Tantau[§]
International Computer Science Institute
1947 Center Street
Berkeley, CA, 94704, USA

October 15, 2004

## Abstract

We study Turing machines that are allowed absolutely no space overhead. The only work space the machines have, beyond the fixed amount of memory implicit in their finite-state control, is that which they can create by cannibalizing the input bits' own space. This model more closely reflects the fixed-sized memory of real computers than does the standard complexity-theoretic model of linear space.

Though some context-sensitive languages cannot be accepted by such machines, we show that all context-free languages can be accepted nondeterministically in polynomial time with absolutely no space overhead, and that all deterministic context-free languages can be accepted deterministically in polynomial time with absolutely no space overhead.

*Keywords.* Overhead-free computation, cannibalistic computation, context-free languages, CFL, deterministic context-free languages, DCFL, linear space, in-place algorithms, space overhead, space reuse, two-stack automata, restarting automata, RRW-automata, editing Turing machines, DLINSPACE.

---

[*]This revises and extends URCS technical report TR-2002-779 and the authors' DLT '03 paper [HMT03], and appears also as URCS technical report TR-2004-844.

[†]Supported in part by grant NSF-CCF-042676.

[‡]mukherji@cs.rochester.edu. Supported in part by grant NSF-IIS-0328849.

[§]tantau@icsi.berkeley.edu. Supported in part by an Erwin-Stephan-Prize grant from Technische Universität Berlin, and a postdoctoral research fellowship grant from the Deutscher Akademischer Austauschdienst (DAAD). Work done in part while working at the Technische Universität Berlin and while visiting the University of Rochester.

# 1 Introduction

Perhaps the most central goal of complexity theory is to gain an understanding of which problems can be solved *realistically* using a computer. Since what resources are deemed "realistic" depends on context, different resource bounds on various models have been studied. For example, deterministic linear space is a possible formalization of the limited memory of computers. Unfortunately, the standard complexity-theoretic formalizations may be too "rough" in realistic contexts, since most have hidden constants tucked away in their definitions. Polynomial-time algorithms with a time bound of $n^{100}$ and linear-space algorithms that need one gigabyte of extra memory per input bit will typically be unhelpful from a practical point of view.

In this paper we study a model that we believe more realistically captures the fixed-sized memory of computers. We use deterministic and nondeterministic one-tape Turing machines that may both read and write their tape. Crucially, we require that the machine may write only on tape cells that are nonempty at the beginning of the computation, when the tape is initialized with the input string (with an unwritable left endmarker $\vdash$ to the input's immediate left and an unwritable right endmarker $\dashv$ to the input's immediate right). The head may move neither left of the left endmarker nor right of the right endmarker. All (and only) words over the input alphabet, which is typically $\{0, 1\}$, are allowed as input strings. Also, crucially, we require that the machine may write only symbols drawn from the *input* alphabet.

In our model of overhead-free computation no auxiliary space is available, but the machine *can* attempt to create work space by "cannibalizing" the space occupied by its input. However, the price of doing so is that the overwritten parts will potentially be lost, unless stored elsewhere via overwriting other parts of the input or unless stored in the machine's finite-state control. The machine is not allowed to cheat by using an enriched tape alphabet. Allowing such cheating would transform our model into one accepting exactly the linear-space languages.

The restrictions we impose on overhead-free machines have quite dramatic consequences on the design of (overhead-free) algorithms. For example, they rob us of the ability to place arbitrary "marker symbols" on the tape. To appreciate how sorely one can miss marker symbols, we invite the reader to try to find an overhead-free algorithm for accepting (even nondeterministically) the set of bit strings that are palindromes. An even more challenging (and as yet unsolved) task is to find an overhead-free algorithm for the innocent-looking language $\{ww \mid w \in \{0, 1\}^*\}$.

We will use notational shortcuts for the four language classes this paper introduces. The class of languages accepted by deterministic overhead-free machines will be denoted DOF, and its nondeterministic counterpart will be denoted NOF. Although these classes "realistically" limit the *space* resources, the underlying machines can still potentially run exponentially long before they decide whether to accept. We will also study which languages can be accepted *time efficiently* by overhead-free machines, that is, in polynomial time. Let $\text{DOF}_{\text{poly}}$ denote the class of those languages in DOF that are accepted by deterministic overhead-free machines running in polynomial time, and define $\text{NOF}_{\text{poly}}$ analogously.

Although deterministic overhead-free computation is a natural model, its nondeterministic counterpart might appear to be of only theoretical interest. After all, nondeterministic computations are hardly "realistic" even if they are overhead-free. However, nondeterministic computations are useful in understanding the inherent limitations of overhead-free computation. An example of such a limitation is the fact that some context-sensitive languages cannot be accepted by overhead-free machines—*not even by nondeterministic ones*.

Previous work on machines with limited alphabet size mostly concerned the *limitations* of such machines. For example, machines—called linear bounded automata with bounded alphabet size—somewhat similar to overhead-free machines have been studied in a note by Feldman and Owings [FO73]. The work of Feldman and Owings implies that DOF is a proper subset of DCSL, the class of all deterministic context-sensitive languages. The work of Seiferas [Sei77] implies that NOF is a proper subset of CSL, the class of all context-sensitive languages.

In this paper we are mostly concerned with the *power* of overhead-free machines. We show that overhead-free machines can simulate certain types of restart automata [JMPV95, Ott03]. As a corollary we obtain the claim that all deterministic context-free languages are contained even in the most restrictive of our four classes, namely $DOF_{poly}$. We present an algorithm that establishes that all context-free languages are contained in $NOF_{poly}$. As additional indicators of the power of overhead-free computation, we point out that $DOF_{poly}$ contains non-context-free sets and that DOF even contains PSPACE-complete sets.

We show that our model is equivalent to other natural models. First, overhead-free machines are exactly as powerful as two-stack machines that may push only input-alphabet symbols and that may never push more symbols than they have popped. This observation allows us to give simple algorithms for accepting numerous languages in an overhead-free way. Second, overhead-free machines are exactly as powerful as editing Turing machines if we enforce appropriate restrictions on the tape size and content. This characterization of overhead-free computation is crucial for our proof that all context-free languages are in $NOF_{poly}$.

**Related Work.** A functional variant of overhead-free computation has strong roots in the literature. There is a large body of work, dating back decades and active through the present day, on *in situ* or "in-place" algorithms. That work is, loosely speaking, interested not in accepting *languages* but rather in implementing, with almost no overhead, *transformations*. As just as few examples of the extensive literature on in-place algorithms, we mention [Dij82, DvG82, Fis92, KP99, GKP00, ALS03].

This paper's focus is on a restricted type of one-tape linear-*space* Turing machine. Those interested in the theory of one-tape linear-*time* Turing machines may find Tadaki, Yamakami, and Lin [TYL03, TYL04], and the references therein, a useful literature starting point. The interesting independent work of Csuhaj-Varju, Ibarra, and Vaszil ([CIV], see also [Iba]) relates the nature of the workspace to the richness of the input seen so far. Their work is in the quite different context of membrane computing, which studies "biomolecular computing devices working in a distributed and parallel manner inspired by the functioning of the living cell ... [and based on] a hierarchically embedded structure of membranes" [CIV]. Informally put, they study membrane models in which the (separate from the input) workspace may at each moment in the computation use just the multiset of "objects" that have already entered the system. They show that, depending on whether membrane rules are applied sequentially or with "maximum parallelness," the power of such membrane systems ranges from being a strict subset of the languages accepted by one-way logspace Turing machines to being exactly the set of all context-sensitive languages. To help understand these membrane systems, they introduce a new model of nondeterministic Turing machine—called 1-way $S(n)$-space-bounded machines—in which, at every moment in the (accepting) computation, the number of nonblank characters on the (separate) worktape(s) is limited not by $S$(input size) but rather is bounded by $S$(current position of the one-way input tape head) [CIV]. Their model differs from our model in multiple ways: They allow their machines to use an arbitrary, richer alphabet on the worktape; they have separate input and work tapes; and their input is read in a one-way fashion.

**Organization.** This paper is organized as follows. In Section 2 we review the basic concepts and define the classes DOF, NOF, DOF$_{\text{poly}}$, and NOF$_{\text{poly}}$. In Section 3 we demonstrate the power of overhead-free computation. In Section 4 we discuss the limitations of overhead-free computation.

## 2 Definitions and Review of Basic Concepts

In this section we first review some basic concepts that will be needed in later sections. We then define the four models of overhead-freeness studied in this paper.

For each finite set $A$, $\|A\|$ denotes the cardinality of $A$. For any integer $n$, $|n|$ denotes the absolute value of $n$. For any string $x \in \Sigma^*$, $|x|$ denotes the length of $x$. Given two alphabets $\Sigma$ and $\Gamma$, a *homomorphism* is a mapping $h\colon \Sigma \to \Gamma^*$. A homomorphism is *isometric* if all words in the range of $h$ have the same length. A homomorphism is extended to words by $h(a_1 \cdots a_n) := h(a_1) \cdots h(a_n)$ and to languages by $h(L) := \{h(w) \mid w \in L\}$. Its *inverse application* is the set $h^{-1}(L) := \{w \mid h(w) \in L\}$. For a machine $M$ let $L(M)$ denote the language accepted by $M$.

Let DLINSPACE denote $\bigcup_{k>0} \text{DSPACE}[kn]$ and let NLINSPACE denote $\bigcup_{k>0} \text{NSPACE}[kn]$. DCFL denotes the class of all deterministic context-free languages (see [HU79]). CFL denotes the class of all context-free languages. CSL denotes the class of all context-sensitive languages. It is well known that CSL = NLINSPACE. The class DCSL (the "deterministic context-sensitive languages") is by definition DLINSPACE. It is not hard to see that DLINSPACE (NLINSPACE) contains the languages accepted by deterministic (nondeterministic) one-tape Turing machines that write on only the cells occupied by the input—but in the model (not ours) in which machines *are* allowed to write arbitrary symbols of a possibly large tape alphabet.

We introduce four complexity classes to capture overhead-free computation. For a rigorous definition we must tackle one technical issue: The notion of overhead-freeness is sensible only if languages "carry around" their underlying alphabet. Normally, the difference between, say, the language $A = \{1^p \mid p \text{ is prime}\}$ taken over the input alphabet $\{1\}$ and the same language $A$ taken over the input alphabet $\{0, 1\}$ is irrelevant in complexity theory, since we can enlarge (in the model—not ours—where the input tape is separate from the work tape) the work tape's alphabet in both cases. In contrast, for overhead-freeness it makes a difference whether the input alphabet is unary or binary, since a unary input alphabet robs us of the possibility of interestingly writing *anything* onto the tape—see Theorem 4.1. Thus, from a formal point of view we consider our complexity classes to contain tuples $(L, \Sigma)$ consisting of a language $L$ and an alphabet $\Sigma$ such that $L \subseteq \Sigma^*$.

For the following definition, recall that we called a machine *overhead-free* if it writes on only those cells that were initially filled with the input and if it writes only symbols drawn from the input alphabet.

**Definition 2.1** *A pair $(L, \Sigma)$ is in the class* DOF *if $L$ is accepted by a deterministic overhead-free machine with input alphabet $\Sigma$. A pair $(L, \Sigma)$ is in* $\text{DOF}_{\text{poly}}$ *if $L$ is accepted by a deterministic overhead-free, polynomial-time machine with input alphabet $\Sigma$. The counterparts to* DOF *and* $\text{DOF}_{\text{poly}}$ *defined in terms of nondeterministic machines are denoted* NOF *and* $\text{NOF}_{\text{poly}}$.

Since differing input alphabets are mainly a technical subtlety, in the following we will speak just of $L$ when the alphabet $\Sigma$ is clear from context.

## 3 The Power of Overhead-Free Computation

In this section we explore the power of overhead-free computation. We start with an explicit example of a non-context-free set that is nonetheless in the smallest of our classes, namely $\text{DOF}_{\text{poly}}$. We then show how overhead-free computation can be characterized by other computational models, namely overhead-free two-stack automata, restarting automata, and editing Turing machines. Building on these characterizations, we relate context-free language classes to overhead-free computation. At the end of the section we show that $\text{DOF}_{\text{poly}}$ contains a P-complete set, that $\text{NOF}_{\text{poly}}$ contains an NP-complete set, and that DOF and NOF both contain PSPACE-complete sets.

### 3.1 A First Example

As an introductory example, we show that the language $A := \{0^n 1^n 0^n \mid n \geq 1\}$ is in $\text{DOF}_{\text{poly}}$. Since this set is not context-free, $\text{DOF}_{\text{poly}}$ contains non-context-free sets.

**Theorem 3.1** *There is a set in* $\text{DOF}_{\text{poly}}$ *that is not context-free.*

*Proof.* We show $A = \{0^n 1^n 0^n \mid n \geq 1\} \in \text{DOF}_{\text{poly}}$ via a machine $M$. On input $w \in \{0, 1\}^*$, using a left-to-right sweep, $M$ first ensures that the input is of the form $0^+1^+0^+$. During the following computation the tape's content will always be of the form $0^*1^*0^*1^*$. The machine now loops through the following instructions: Accept if the tape contains a string of the form $0101^*$. Otherwise, move back to the left end. In the first 0's block replace the last 0 by a 1, which enlarges the following 1's block by one 1. In this block replace the last two 1's by 0's, which enlarges the following 0's block by two 0's. In this block replace the last three 0's by 1's. If at any point you inadvertently hit the right endmarker or have a missing block, reject. Otherwise, return to the left end and repeat.

Clearly, the computation is overhead-free and runs in quadratic time. Furthermore, the input word will be accepted exactly if it is of the prescribed form. □

## 3.2 Two-Stack Automata and Overhead-Free Computation

In this subsection we give the first characterization of overhead-free computation in terms of another model. We show that two-stack automata of appropriately bounded total stack height are exactly as powerful as overhead-free machines. This observation is useful because it is sometimes easier to describe algorithms for two-stack machines than for overhead-free machines.

**Definition 3.2** *A* two-stack automaton *has two stacks whose bottoms are indicated by a special bottom symbol. Initially, the first stack is filled with an* input word $w \in \Sigma^*$ *such that the first letter is topmost. In each step, depending on the two top symbols of the stacks and depending on its current internal state, the automaton changes its internal state and may choose to pop a symbol from one of the stacks, to leave the stacks unchanged, or to push a symbol from a tape alphabet $\Gamma \supseteq \Sigma$ onto one of the stacks. When the automaton enters a special accepting state, the word is accepted. The set of all words accepted by a two-stack automaton $M$ is denoted $L(M)$. A two-stack automaton is* overhead-free *if (a) $\Sigma = \Gamma$ and (b) during every computation the sum of the heights (not counting the special bottom symbols) of the two stacks never exceeds the length of the input word.*

As is well known, two-stack automata are exactly as powerful as general Turing machines if there is no restriction on the stack heights. Overhead-free two-stack automata are obviously less powerful (their configuration space is exponentially bounded), but can still accept interesting languages such as, for instance, the set of palindromes: An automaton for this language works as follows: It accepts if the input is $\epsilon$. Otherwise, it pops the top symbol from the first stack and stores it in its internal state. Then it pops all symbols from the first stack and pushes them onto the second stack. When the first stack is empty, the automaton accepts if the second stack is empty, and otherwise pops the top symbol from the second stack and compares it with the stored symbol. If these are not the same, the automaton rejects. If they are the same, then the automaton pops all remaining symbols from the second stack and pushes them onto the first one. It repeats this procedure until both stacks are empty, at which point it accepts.

**Theorem 3.3** *If $L \subseteq \Sigma^*$ and $\|\Sigma\| \geq 2$, then all of the following hold:*

1. *$L \in \text{DOF}$ if and only if there exists a deterministic overhead-free two-stack automaton $M$ with $L(M) = L$.*

2. $L \in$ NOF *if and only if there exists a nondeterministic overhead-free two-stack automaton $M$ with $L(M) = L$.*

3. $L \in \text{DOF}_{\text{poly}}$ *if and only if there exists a deterministic polynomial-time overhead-free two-stack automaton $M$ with $L(M) = L$.*

4. $L \in \text{NOF}_{\text{poly}}$ *if and only if there exists a nondeterministic polynomial-time overhead-free two-stack automaton $M$ with $L(M) = L$.*

*Proof.* For the first direction of part 1 of the theorem, let $L \in$ DOF via a deterministic overhead-free machine $M'$. In order to simulate $M'$ by an overhead-free two-stack automaton $M$, we use the two stacks as follows: All tape symbols under or following the head of $M'$ are on the first stack of $M$ with the symbol under the head topmost on that stack; all tape symbols before the head of $M'$ are on the second stack of $M$ with the symbol just to the left of the head topmost. The different operations of $M'$ can easily be simulated in such a way that this invariant is maintained. For instance, a right move of $M'$ corresponds to popping the topmost symbol from the first stack and pushing to the second stack whatever symbol $M'$ wrote to the cell it was moving off of.

For the other direction of part 1, let $M$ be an overhead-free two-stack automaton. It is simulated by an overhead-free machine $M'$ as follows. The tape of $M'$ has three parts. The first part, which starts after the left endmarker, is the contents of the second stack with the topmost symbol at the right. This part is followed by a word of the form $10^*1$ (0 and 1 being distinct symbols in $\Sigma$), by the single symbol 1, or by the empty word $\epsilon$. At all times, $M'$ keeps track (in its internal state) of which of these three forms the middle part currently has. Initially the middle part is $\epsilon$. The final part fills the tape up to the right endmarker and stores the first stack with its topmost symbol to the left.

The head of the machine $M'$ is always inside the middle part or on one of the symbols adjacent to it. $M'$ simulates the operations of $M$ as follows: Popping a symbol from the first stack corresponds to shifting the middle part one link along the chain $\epsilon \to 1 \to 11 \to 101 \to 1001 \to \ldots$, and is done by expanding the middle part one space rightward. Similarly, popping a symbol from the second stack means the same expansion, but expanding the middle part one space leftward. Pushing a symbol to the first stack is a contraction along this chain—freeing up the rightmost space of the middle part and overwriting this freed position by the desired symbol. Pushing a symbol to the second stack is implemented similarly.

Consider a simulation of the machine $M$ by the machine $M'$. By assumption, at all times the total height of $M$'s two stacks never exceeds the input length. So during the simulation the size of the middle part never drops below zero. This shows that $M'$ will reach exactly the same end state as $M$ does and that a word is accepted by $M$ if and only if that word is accepted by $M'$.

For part 2 of the theorem, the constructions work exactly the same way. For parts 3 and 4, note that in the first direction of part 1 above every step of the overhead-free machine $M'$ can be simulated in a constant number of steps by the overhead-free two-stack automaton $M$, and that in the second direction of part 1 above every step of $M$ can be simulated in a linear number of steps by $M'$. □

The above theorem does not hold for unary alphabets. Overhead-free machines over unary alphabets clearly can only accept regular languages, but the language $\{a^{2^n} \mid n \in \mathbb{N}\}$ can be accepted by a unary two-stack automaton: For every two symbols on the first stack, push one symbol onto the second stack; then for every two symbols on the second stack, push one symbol onto the first stack; and so on. If at the start of such a stack-to-stack shifting phase there is a single $a$ on the stack we are shifting from, then accept. If at the end of a stack-to-stack shifting phase there is an orphaned $a$ on the stack we are shifting from, then reject.

In the next subsections we will see that overhead-free Turing machines are surprisingly powerful. By Theorem 3.3, this also tells us something about the power of overhead-free two-stack automata. For example, by the results of the following subsections overhead-free two-stack automata can deterministically accept all deterministic context-free languages and can nondeterministically accept all context-free languages.

## 3.3 Restarting Automata and Overhead-Free Computation

The next model we compare overhead-free computation to is that of restart automata. We show that a special type of restart automata, namely RRW-automata, can be simulated using overhead-free machines. Since RRW-automata are powerful enough to accept all deterministic context-free languages, all deterministic context-free languages can also be accepted in an overhead-free way.

Many variants of restarting automata have been studied; see [Ott03] for an overview. We concentrate on RRW-automata since they appear to be the most powerful restarting automata that can be simulated in an overhead-free way.

**Definition 3.4 ([JMPV98])** *An* RRW-automaton *is a one-tape machine whose tape is initialized with an input word $w \in \Sigma^*$, delimited by left and right endmarkers. It "sees" the input through a fixed-size window. It can perform four kinds of operations:*

1. *A* move-right step, *which moves the window one symbol to the right, possibly changes the internal state, but leaves the tape contents unmodified.*

2. *A* rewrite step, *which causes the contents of the window to be replaced by a shorter string drawn from $\Sigma^*$, though the endmarkers must be reproduced if a replacement takes place at an end. The tape is shortened when such a step is performed.*

3. *A* restart step, *which causes the head to return to the left end of the tape, and the machine to reenter the initial state.*

4. *An* accept step, *which causes the machine to accept.*

*Moreover, between any two executions of a restart step there must be at least one rewrite step, and between any two rewrite steps there must be at least one restart step. (The effect of this is that the two types of steps—ignoring other types—alternate.) The class of languages accepted by nondeterministic RRW-automata is denoted $\mathcal{L}(\mathrm{RRW})$. The deterministic version is denoted $\mathcal{L}(\mathrm{det\text{-}RRW})$.*

As an example, a simple RRW-automaton that accepts the language $\{a^n b^n \mid n \geq 1\}$ works as follows. It accepts if its tape contains precisely the string *ab*. Otherwise, it scans its input until it finds an occurrence of *aabb*, which it replaces with *ab* and restarts.

**Theorem 3.5** $\mathcal{L}(\mathrm{det\text{-}RRW}) \subseteq \mathrm{DOF}_{\mathrm{poly}}$ *and* $\mathcal{L}(\mathrm{RRW}) \subseteq \mathrm{NOF}_{\mathrm{poly}}$.

*Proof.* Let $A \in \mathcal{L}(\mathrm{det\text{-}RRW})$ via an RRW-automaton $M$. By Theorem 3.3 it suffices to show that $M$ can be simulated by a deterministic polynomial-time overhead-free two-stack automaton $M'$.

Let $w \in \Sigma^*$ be an input word on the first stack (first letter uppermost). The machine $M'$ uses its finite-state memory to keep a record of the substring $M$ "sees" through its window. $M'$ constructs this record initially by popping as many symbols from the first stack as $M$'s window size dictates. It then checks what operation $M$ would perform on "seeing" that substring, and simulates it as follows.

If $M$ performs a move-right step, $M'$ pushes the leftmost symbol of the window onto the second stack, pops the next symbol from the first stack, and adjusts its record of the window contents accordingly. If at any point there are insufficient characters left on the first stack to fill the window, $M'$ behaves as though a right endmarker had been observed.

If $M$ performs a rewrite step, $M'$ simply switches its record of the window contents to the new string and then pops as many symbols from the first stack as are needed to fill up the window (with insufficient characters indicating a right endmarker).

If $M$ performs a restart step, $M'$ pushes the stored window contents back onto the first stack and then pops each element in turn from the second stack and pushes it back onto the first stack.

Finally, if $M$ accepts, $M'$ accepts also.

There is a quadratic time bound on the computation length since after each restart the sum of the stacks' sizes must decrease by at least one.

The construction works exactly the same way for nondeterministic automata. □

## 3.4 Editing Turing Machines and Overhead-Free Computation

We now introduce another computational model that is closely related to overhead-free computation. We call this model the "editing Turing machine." In this subsection we show that editing Turing machines can be simulated by overhead-free machines if their tapes have "bounded weight," a concept we will define presently. Apart from being a very natural extension of the standard Turing machine model, the concept of editing Turing machines allows us later on to give an elegant proof that all context-free languages can be accepted in an overhead-free way.

**Definition 3.6** *An* editing Turing machine *is a one-tape Turing machine whose tape is initialized with an input word, bordered by endmarkers. We allow two additional operations apart from the usual ones:*

1. *An* insert *operation, which inserts a new cell at the current head position (so, for example, the cell under the head position before the* insert *will be, after the* insert*, just to the right of the inserted cell), thus increasing the number of cells between the endmarkers by one. The* insert *operation is not permitted on the left endmarker.*

2. *A* delete *operation, which removes the current cell from the tape, after which the head sits on the cell formerly to the right of the deleted cell, thus decreasing the number of cells between the endmarkers by one. The* delete *operation is not permitted on either endmarker.*

**Definition 3.7** *Let $M$ be a one-tape Turing machine with input alphabet $\Sigma$ and tape alphabet $\Gamma$. For a given configuration let $w_1 \cdots w_\ell$ be the contents of the tape cells between the endmarkers and let $p$ denote the head position with $p = 0$ when the head is over the left endmarker. The* weight *of the configuration is defined as $\sum_{i=1}^{\ell} \mu_i$, where*

$$\mu_i = \begin{cases} 1 & \text{if } w_i \in \Sigma \text{ or } i = p, \\ 1 + 3\log_2(|p - i|) & \text{otherwise.} \end{cases}$$

The weight of a configuration clearly decreases every time a *delete* operation is performed. The weight function for noninput symbols is clearly nondecreasing in their distance from the head, and indeed slowly increases the farther they are from the head. Thus, intuitively, to keep the configurations' weights low during a computation, we must try to keep the computation as "local" as possible: Very loosely speaking, the head can be allowed to venture far from noninput symbols only if many other symbols have already been deleted.

The weight of any configuration of an overhead-free machine (recall that such machines have no noninput symbols or *insert* or *delete* operations) is constant: It is always equal to the input length. In particular, the weight is bounded above by the input length. The following theorem establishes that the converse is also true: If an editing Turing machine has the property that for strings in the language there is always at least one accepting path such that at every step of that path the weight of its configuration does not exceed the length of the original input, then it can be simulated by an overhead-free machine. In fact, even if exceeding the input's length by a—global for the machine—constant is allowed, the simulation can still be performed.

**Theorem 3.8** *Let $L \subseteq \Sigma^*$ and $\|\Sigma\| \geq 2$. Suppose $M$ is an editing Turing machine with $L = L(M)$, and that there is a constant $c$ such that, for each $x \in L$, for at least one accepting path of $M$ on input $x$ every configuration on that path is bounded in weight by $|x| + c$. Then the following implications hold: (a) If $M$ is deterministic, then $L \in \text{DOF}$; (b) if $M$ is nondeterministic, then $L \in \text{NOF}$; (c) if $M$ is deterministic and polynomially time-bounded, then $L \in \text{DOF}_{\text{poly}}$; (d) if $M$ is nondeterministic and polynomially time-bounded, then $L \in \text{NOF}_{\text{poly}}$.*

*Proof.* We prove only the fourth claim; the proofs of the other claims are similar.

We assume $\{0,1\} \subseteq \Sigma$. We construct an overhead-free machine $M'$ that simulates $M$. We first describe how this can be done for the case $\Sigma = \Gamma$. Later on, we describe how additional symbols can be incorporated.

For the case $\Sigma = \Gamma$, we need only describe how to implement the *insert* and *delete* operations. The idea is to introduce a flexible *free space area*, which we insert at the current head position (and will drag around with us). As in the proof of Theorem 3.3, the free space area is a string, either of the form $10^*1$, or just a 1, or the empty string $\epsilon$, and in our finite-state control we at all times keep track of which of these cases currently holds. The head always remains inside the area or on a cell adjacent to it. When $M$ performs a left or right move, the free space area is moved one symbol to the left or right. When $M$ performs a *delete* operation, the free space area grows by one symbol; when $M$ performs an *insert* operation, it shrinks by one symbol.

In detail, a right move is implemented as follows. There are three cases. The first case is if the free space area is currently $\epsilon$. In this case the head simply moves one space to the right. The second case is if the free space area is currently 1. In this case the symbol in the cell just to the right of the free space area is exchanged with the 1 of our free space area, with our head ending up one cell to the right of where it started, namely, over the new free space cell. The third case is if the free space area is currently of the form $10^*1$. Briefly put, we will rewrite the $10^k 1\alpha$ to become $\alpha 10^k 1$, where $\alpha$ is the cell to the right of the free space area. In full detail: In this third case, the symbol in the cell just to the right of the free space area is read and recorded in our finite state control. We overwrite that just-read cell with a 1, which will be the right-end 1 of our new free space area. Then our head will move left over the (zero or more) 0 symbols that are to the left of this just-written 1 until the head reaches another 1. That 1 is then overwritten by the symbol we recorded in our internal state, and the cell just to the right of that is overwritten by a 1, which will be the left end of our new free space area.

Left moves, are implemented similarly. The *insert* and *delete* operations are implemented in the natural way.

So far we have not handled the "plus some constant" in the theorem statement. This can be done as follows. Note that we can, for any fixed constant $c$, handle our tape as having up to $c$ extra cells that we will imagine as being "located" between the free space area and the cell just to its right but actually stored in our finite control. We also record in the finite control which, if any, of these virtual cells the virtual tape head is over. This handles the "plus some constant" issue. (Note that this extra space cannot be "located" between the tape's rightmost cell and the right endmarker, because we cannot uniquely mark a position on tape in our model, and thus would not be able to return to the head position after "expanding" the tape if we did that).

Let us now consider how to simulate an editing Turing machine $M$ for which $\Sigma \subsetneq \Gamma$. In this case, we use part of the free space to keep track of any symbols from $\Gamma - \Sigma$ that are on $M$'s tape. We will ensure that the size of the part we use for bookkeeping never exceeds the size of the free space area.

The unused part (rightmost part) of the free space area will still consist just of 0's, separated from the bookkeeping part by a 1. We will ensure that the coding of each bookkeeping record over the alphabet $\{0,1\}$ is self-delimiting. With these precautions, we will still be able to simulate *insert*s, *delete*s, and regular moves—even when these involve symbols from $\Gamma - \Sigma$.

In detail, let $g\colon (\Gamma - \Sigma) \to \{0,1\}^k$ be an injective mapping, where $k = \lceil \log_2 \|\Gamma - \Sigma\| \rceil$. Let $h$ map any natural number $n$ to a self-delimiting bit string of length $2\lfloor \log_2(n+1) \rfloor + 4$. Let us say this is accomplished by encoding natural number $n$ as the string $s(n)$, the $(n+1)$st string of $\{0,1\}^*$ in lexicographic order, and then, further, re-encoding the bits of $s(n)$—call them $s_1 s_2 \cdots s_m$—as the string $11 s_1 0 s_2 0 \cdots s_m 011$.

A bookkeeping record is then a bit string of the form

$$f_1 f_2 f_3 g(\gamma) \sigma h(n)$$

for $f_1, f_2, f_3 \in \{0,1\}$, $\gamma \in \Gamma - \Sigma$, $\sigma \in \{0,1\}$, and $n \in \mathbb{N}$. It is interpreted as follows: The first three bits are flags. $f_1$ and $f_2$ respectively signal the leftmost and rightmost bookkeeping records on $M'$'s tape, so that the ends of the bookkeeping portion of the free space can be identified. $f_3$ is a flag that allows $M'$ to identify the last record updated when shifting the bookkeeping records around as described below. The next $k$ bits represent $\gamma$, a symbol from $\Gamma - \Sigma$ that has been written to tape (if $\|\Gamma - \Sigma\| = 1$, the additional character is mapped to $\epsilon$). The remaining bits indicate how many cells distant that symbol is from the current head position *in the simulation of $M$*. These distances are negative for symbols from $\Gamma - \Sigma$ to the left of $M$'s head; positive for those to the right. The number is stored as $\sigma h(n)$, where $\sigma$ is a bit indicating the sign of the number to follow, which is stored as a self-delimiting string.

Every time the machine $M$ overwrites a symbol $\delta \in \Sigma$ with a symbol $\gamma \in \Gamma - \Sigma$, the machine $M'$ instead deletes $\delta$ (this increases the size of the free space area) and adds a new bookkeeping record. This record stores the symbol $\gamma$ and the distance of $\gamma$ from $M$'s current head position as described above. Upon creation, this distance is obviously zero. However, whenever $M$'s head moves, this distance is updated appropriately: When it moves left, all distances are increased by one; when it moves right, the distances are decremented. In each update step, the sizes of the different bookkeeping records may increase or decrease, in which case they are shifted around inside the free space area to keep the bookkeeping part compacted.

Bookkeeping records are also created when $M$ inserts a symbol from $\Gamma - \Sigma$ using the insert operation. On the other hand, they can be deleted if $M$ performs a delete operation on a symbol in $\Gamma - \Sigma$, or if $M$ overwrites a symbol from $\Gamma - \Sigma$ with one from $\Sigma$. In each case, the distances must be updated appropriately.

Finally, if at any point $M'$ runs out of space (beyond the uses mentioned above of constant extra memory in the finite control), it immediately rejects.

It remains to show that, if $x \in L$, there is at least one accepting path in $M'$. In other words, we have to show that there is at least one accepting path on which the size of the bookkeeping records never exceeds that of the free space area. We know that if $x \in L$, there is at least one accepting path on which the weight of $M$'s configuration never exceeds the input length (plus some constant, which we can take care of using our finite-state memory as described earlier). So on such a path we have $1 + 3\log_2(|p - i|)$ bits for storing a bookkeeping record for the symbol at position $i$ when $M$'s head is at position $p \neq i$. The size of a bookkeeping record is $3 + k + 1 + 2\lfloor\log_2(|p - i| + 1)\rfloor + 4$. Now with $p \neq i$, $\lfloor\log_2(|p - i| + 1)\rfloor \leq \log_2(|p - i|) + 1$, so $1 + 3\log_2(|p - i|)$ bits is certainly sufficient whenever $\log_2(|p - i|) > k + 10$. However, there can be at most one character at each distance $p - i$ from the head position. Thus there will be at most a constant number of bookkeeping records that require additional space (i.e., for which $\log_2(|p - i|) \leq k + 10$), and each one requires only a constant amount (at most $k + 10$ bits). This constant extra space can be provided by the finite control. □

Though the "for each $x \in L$, for at least one accepting path of $M$ on input $x$ every configuration on that path is bounded in weight by $|x| + c$" of Theorem 3.8 may seem a bit unnatural to some, we mention in passing that the arguably more natural variant of Theorem 3.8 in which that is replaced by "the weight of every configuration reached during executions of $M$ is bounded by the length of the input plus $c$" follows easily from Theorem 3.8. Furthermore, the greater flexibility in the formulation of Theorem 3.8 will be central in allowing us to invoke Theorem 3.8 during the proof that CFL $\subseteq$ NOF$_{\text{poly}}$ (Theorem 3.15).

## 3.5 Formal Language Classes and Overhead-Free Computation

Formal language classes, such as the class of regular languages or the class of context-free languages, are related to overhead-free computation in different ways. For example, (non)deterministic overhead-free computation is powerful enough to decide all (non)deterministic context-free languages. The proofs of these theorems make heavy use of the results we obtained earlier regarding restart automata and editing Turing machines.

The regular languages are clearly (via machines that move their heads steadily to the right and never write at all) even in DOF$_{\text{linear}}$.

**Theorem 3.9** *All regular languages are in* DOF$_{\text{poly}}$.

The next formal language class we study is the class of deterministic context-free languages. These languages are in DOF$_{\text{poly}}$. The proof of this result is based on the following somewhat surprising advance.

**Theorem 3.10 ([JMPV99])** *All deterministic context-free languages are in* $\mathcal{L}(\text{det-RRW})$.

From Theorems 3.5 and 3.10, we have the following.

**Corollary 3.11** *All deterministic context-free languages are in* DOF$_{\text{poly}}$.

There is another formal language class, namely the Church–Rosser congruential languages, for which we can prove containment in DOF$_{\text{poly}}$ by invoking restarting automata. See [MNO88, Nar84] for details on the definition and properties of these languages.

**Definition 3.12 ([MNO88, Nar84])** *A language $L \subseteq \Sigma^*$ is a Church–Rosser congruential language (CRCL) if there exists a finite, length-reducing, confluent string-rewriting system $R$ on $\Sigma$ and a finite set of irreducible strings $\{w_1, \ldots, w_n\}$ such that $L = \bigcup_{i=1}^n [w_i]_R$.*

**Theorem 3.13 ([NO00a, NO00b])** $\mathrm{CRCL} \subseteq \mathcal{L}(\text{det-RRW})$.

From Theorems 3.5 and 3.13, we have the following.

**Corollary 3.14** $\mathrm{CRCL} \subseteq \mathrm{DOF}_{\text{poly}}$.

We now prove that all context-free languages are in $\mathrm{NOF}_{\text{poly}}$. Our proof makes use of the characterization of overhead-free computation in terms of editing Turing machines (Theorem 3.8).

**Theorem 3.15** $\mathrm{CFL} \subseteq \mathrm{NOF}_{\text{poly}}$.

*Proof.* As is common, by Chomsky Normal Form we mean that each production either turns one nonterminal into exactly two nonterminals, or turns one nonterminal into exactly one terminal. It is well-known that all CFLs that do not contain $\epsilon$ have Chomsky Normal Form grammars.

Let $L \in \mathrm{CFL}$. We will henceforward assume $\epsilon \notin L$. (If $\epsilon \in L$, use the following construction to build an $\mathrm{NOF}_{\text{poly}}$ machine for $L - \{\epsilon\}$, and then patch the machine to also accept $\epsilon$.) Let $G = (N, T, S, P)$ be a grammar ($N$ the nonterminals, $T$ the terminals, $S$ the start symbol, and $P$ the production set; $N \cap T = \emptyset$) in Chomsky Normal Form that generates $L$. We show that $L$ can be accepted by an editing Turing machine in such a way that the weight of every configuration is bounded by the input length plus some constant.

The editing Turing machine has the input alphabet $T$ and the tape alphabet $N \cup T$. It performs a bottom-up parse of the input word $w$ by inverting in nondeterministically chosen places nondeterministically chosen production rules $l \to r$. That is, starting from the current head position, it searches either to the left or to the right for an occurrence of $r$ on the tape (not necessarily the nearest one—we'll act nondeterministically in both what rule we are trying to reverse and what instance of its right-hand side we act on). Having settled on such, it applies a sequence of (one or two) *delete* operations to get rid of $r$, and then an *insert* operation to insert $l$ instead. When the tape contains only the symbol $S$, our machine accepts. If at any point the simulation runs out of space, it rejects (on the current nondeterministic path).

In light of Theorem 3.8, it remains to show that for every word $w \in L$ there exists an accepting computation of $M$ the weights of whose configurations are all bounded by the input length plus some constant. Let $T = (R, E)$ be a binary parse tree of $w$. The node set $R$ contains all rule *applications* $X \to YZ$ or $X \to a$, with $X, Y, Z \in N$ and $a \in T$, used in the derivation of $w$ (we focus not on rules, but on rule applications; so a given rule may appear in more than one node, namely, if it is applied multiple times in the derivation tree). The edge relation $E$ relates each rule application $r = X \to YZ$ to the rule applications $r_Y$ and $r_Z$ used on $Y$ and $Z$ in the derivation. For each rule application $r \in R$, let $n(r)$ denote the number of leaves in the subtree rooted at $r$ and let $N(r)$ denote the nonterminal symbol on the left-hand side of the rule associated with $r$.

Consider the sequence of nondeterministic choices that causes the parse to proceed as follows: For each nonleaf node $r$ with children $r_Y$ and $r_Z$, these children are parsed one after the other; if $n(r_Y) \geq n(r_Z)$, then $r_Y$ is parsed first; otherwise $r_Z$ is parsed first. This parse ordering ensures that the weight never exceeds the input length. To see this, consider the nonterminal symbols on the tape at any time in the course of the computation. Each of these symbols corresponds to a node $r$ whose subtree has already been parsed, and whose sibling's subtree is in the process of being parsed or, better still, has just been parsed and we are about to implement the rule inversion that combines these two siblings' nonterminals into a single nonterminal. Since $r$'s sibling's subtree is smaller than $r$'s subtree, the head will never be more than $n(r)$ cells away from the symbol $N(r)$.

Thus, every nonterminal $N(r)$ corresponds to the deletion of $n(r)$ terminals (which reduces the weight by $n(r)$) and adds at most $1 + 3\log_2 n(r)$ to the weight. Thus, the contribution of each nonterminal to the total weight is nonpositive, and the weight of every configuration of $M$ is bounded by the input length. □

## 3.6 Complete Problems in Overhead-Free Computation Classes

Our final aim for this section is to show that all overhead-free computation classes contain problems that are complete for classical complexity classes. For the proof, we first show that given any language $L \in \mathrm{DLINSPACE}$ we can find a closely related language $L' \subseteq \{0,1\}^*$ that is accepted by an overhead-free machine.

**Lemma 3.16** *Let $L \in \mathrm{DLINSPACE}$ with $L \subseteq \Sigma^*$. Then there exists an injective isometric homomorphism $h\colon \Sigma \to \{0,1\}^*$ such that for $L' := h(L) \subseteq \{0,1\}^*$ we have $L' \in \mathrm{DOF}$. Similarly, if $L \in \mathrm{NLINSPACE}$, then there exists an injective isometric homomorphism $h$ such that $L' := h(L) \in \mathrm{NOF}$.*

*Proof.* Let $L \in \mathrm{DLINSPACE}$ via a machine $M$ that never writes on any cells other than those already initially occupied by the input. As mentioned earlier, every language in DLINSPACE can be accepted in such a fashion. Let $M$ use the tape alphabet $\Gamma \supseteq \Sigma$, which may be strictly richer than $\Sigma$. Let $k := \lceil \log_2 \|\Gamma\| \rceil$. Then there exists an injective mapping $g\colon \Gamma \to \{0,1\}^k$ that codes every symbol in $\Gamma$ as a binary string of length $k$. Let $h$ be the restriction of $g$ to the domain $\Sigma$.

We now show $h(L) \in \mathrm{DOF}$ via a machine $M'$. This machine always reads $k$ symbols as a block. On input $w$ it first ensures that $w$ consists only of blocks that encode symbols from $\Sigma$. Then $w = g(u_1)g(u_2)\cdots g(u_m)$ for some sequence of $u_i$'s in $\Sigma$. The machine $M'$ returns to the beginning of the input and starts a simulation of what $M$ would do on input $u$. Every time $M$ reads/writes a single symbol $\delta \in \Sigma$, $M'$ reads/writes the block $g(\delta) \in \{0,1\}^k$. This way, whenever the machine $M$ would reach a state $q$ with tape content $u'_1 \cdots u'_m \in \Gamma^*$, the machine $M'$ will similarly reach the state corresponding to $q$ and will at that point have tape content $g(u'_1)\cdots g(u'_m) \in \{0,1\}^*$. If $M$ would accept, $M'$ does, and vice versa.

For the nondeterministic case the simulation works the same way. □

Since DLINSPACE and NLINSPACE are clearly closed under inverse isometric homomorphism application, and DLINSPACE ⊇ DOF and NLINSPACE ⊇ NOF, we have the following.

**Corollary 3.17** *The closure of* DOF *under inverse isometric homomorphism application is* DLINSPACE. *The closure of* NOF *under inverse isometric homomorphism application is* NLINSPACE.

The closure of both DLINSPACE and NLINSPACE under $\leq_\mathrm{m}^\mathrm{log}$-reductions is PSPACE, so we also have the following.

**Corollary 3.18** *The class* DOF *contains a $\leq_\mathrm{m}^\mathrm{log}$-complete set for* PSPACE.

It is not hard to also see the following.

**Corollary 3.19** *The class* DOF$_\mathrm{poly}$ *contains a $\leq_\mathrm{m}^\mathrm{log}$-complete set for* P.

**Corollary 3.20** *The class* NOF$_\mathrm{poly}$ *contains a $\leq_\mathrm{m}^\mathrm{log}$-complete set for* NP.

The fact that powerful sets reduce to DOF does not say that those sets are in DOF themselves. In fact, DOF is a proper subset of PSPACE since by the space hierarchy theorem some PSPACE languages are not in DLINSPACE, let alone DOF.

## 4 Limitations of Overhead-Free Computation

The previous section demonstrated that several interesting languages can be accepted by overhead-free machines. In this section we discuss what cannot be done using overhead-free machines.

We begin with an observation that shows that overhead-free machines on unary alphabets are just as powerless (or powerful, depending on your point of view) as finite automata. Theorems 4.2 and 4.3 then show that there are (non)deterministic context-sensitive language that cannot be accepted by (non)deterministic overhead-free machines. Both results are based on diagonalization techniques. In the rest of the section we discuss whether certain *natural* sets can be accepted in an overhead-free fashion.

The comments made earlier in the paper about language–alphabet pairs are relevant to the following theorem. For example, the "$L \in$ DOF" in the theorem statement really means (in the sense of Definition 2.1) that $(L, \Sigma)$ is in DOF, where $\Sigma$ is a *unary* alphabet. The theorem would break down were one to consider the same language $L$ as being embedded in a richer input alphabet.

**Theorem 4.1** *Let $L$ be a tally set. Then the following are equivalent: (a) $L$ is regular, (b) $L \in$* DOF$_\mathrm{poly}$*, (c) $L \in$* DOF*, (d) $L \in$* NOF$_\mathrm{poly}$*, and (e) $L \in$* NOF.

*Proof.* If $L$ is regular, then by Theorem 3.9 it is in $\text{DOF}_{\text{poly}}$ and so is in DOF, $\text{NOF}_{\text{poly}}$, and NOF. For the other direction, let $L \in \text{NOF}$ via an overhead-free machine $M$. Since the input alphabet is unary, $M$ behaves exactly like a two-way nondeterministic finite automaton and can thus accept only regular sets. □

**Theorem 4.2** $\text{DOF} \subsetneq \text{DLINSPACE}$.

*Proof.* This follows immediately from Corollary 2 of a paper by Feldman and Owings [FO73]. They show that, for every constant $m$, deterministic linear-bounded automata with alphabet size at most $m$ cannot accept all deterministic context-sensitive languages. □

**Theorem 4.3** $\text{NOF} \subsetneq \text{NLINSPACE}$.

*Proof.* Seiferas [Sei77] has shown that for every $m$ there exists a language in NLINSPACE that cannot be accepted by any nondeterministic off-line Turing machine that uses only $m$ different symbols on its tape and uses only as many cells on its work tape as there are symbols in the input. Since any overhead-free machine can be simulated by such an off-line machine by first copying the input to the work tape, we get the claim. □

An alternative proof of Theorem 4.3 can be based on combining Corollary 1 of the paper of Feldman and Owings [FO73] with the Immerman–Szelepcsényi Theorem [Imm88, Sze88]. Corollary 1 of Feldman and Owings states that for each $m$ there is a language whose complement is context-sensitive and cannot be accepted by a nondeterministic linear-bounded automaton whose alphabet size is bounded by $m$. This is an example of the often encountered effect that the Immerman–Szelepcsényi technique can be used to simplify nondeterministic space hierarchy proofs (see [Imm88, Ges93]).

Theorems 4.2 and 4.3 show that overhead-free computation is less powerful than linear-space computation *in principle*. A nice step would be to prove that certain simple, natural context-sensitive languages cannot be accepted in an overhead-free fashion. Our candidate for a context-sensitive language that is not in NOF is $L := \{ww \mid w \in \{0,1\}^*\}$.

Interestingly, though we name $L$ a candidate non-NOF language, it is not hard to show that $L$ is in "2-head-$\text{DOF}_{\text{poly}}$," the analog of $\text{DOF}_{\text{poly}}$ in which the overhead-free machine has two heads. This observation raises the question of how powerful extra heads make our model. First, by a classic result of Hartmanis [Har72], even $\mathcal{O}(1)$-head *finite automata* taken collectively yield the power of logarithmic-space Turing computation. Thus, at least in that different context, additional heads are well known to be a valuable resource. However, using the same argument as in Theorem 4.3, Seiferas' results [Sei77] can be used to show that for every $m$ there exists a context-sensitive language that is not in $m$-head-NOF.

## 5  Conclusion

In this paper we introduced a computational model, namely overhead-free computation, in which Turing machines are allowed to manipulate their input, but may not use any symbols other than those of the input alphabet. This models what a computer can compute by cannibalizing its noncontrol memory, if the input initially fills the whole noncontrol memory. Building on this model we defined the four complexity classes DOF, NOF, $\text{DOF}_{\text{poly}}$, and $\text{NOF}_{\text{poly}}$ and studied how these classes relate to standard formal-language and complexity classes. The most "realistic" of the four classes is $\text{DOF}_{\text{poly}}$, which contains the languages that can be accepted efficiently (that is, in polynomial time) with absolutely no space overhead.

We showed that overhead-free computation is related to other computational models. In particular, overhead-free computation directly corresponds to overhead-free two-stack automata. We showed that overhead-free machines can simulate RRW-automata and editing Turing machines. The smallest of our classes, namely $\text{DOF}_{\text{poly}}$, contains all deterministic context-free languages, and also some non-context-free languages. The class $\text{NOF}_{\text{poly}}$ contains all context-free languages. Curiously, despite the similarity of the statements, the proofs of these last two results are not related.

The relationship between overhead-free computation and formal language classes is interesting. Our results show $\text{DCFL} \subsetneq \text{DOF} \subsetneq \text{DCSL}$ and $\text{CFL} \subsetneq \text{NOF} \subsetneq \text{CSL}$. In other words, overhead-free computation is properly snuggled between the classes of context-free and context-sensitive languages.

**Acknowledgments**  We thank Edith Hemaspaandra, František Mráz, Mitsunori Ogihara, Friedrich Otto, Martin Plátek, and Jonathan Shaw for many helpful comments. We thank Oscar Ibarra for bringing to our attention the work of Csuhaj-Varju, Ibarra, and Vaszil.